\documentclass[12pt,a4paper]{article}
\usepackage{bm}
\usepackage{color}
\usepackage{amsmath}
\usepackage{amsthm}
\usepackage{amssymb}
\usepackage{graphicx}
\usepackage{verbatim}
\usepackage[breaklinks]{hyperref}
\usepackage{multirow}

\newtheorem{theorem}{Theorem}

\newtheorem{claim}{Claim}

\theoremstyle{definition}
\newtheorem{definition}{Definition}

\newcommand{\set}[2]{\ensuremath{ \{ \, #1 \mid #2 \, \} }}

\renewcommand{\emptyset}{\varnothing}
\renewcommand{\epsilon}{\varepsilon}

\begin{document}
\sloppy

\title{Exact descriptional complexity of determinization of input-driven pushdown automata}
\author{Olga Martynova\thanks{
	Department of Mathematics and Computer Science,
	St.~Petersburg State University,
	7/9 Universitetskaya nab., Saint Petersburg 199034, Russia,
	\texttt{olga22mart@gmail.com}.
}}
\maketitle

\begin{abstract}
The number of states and stack symbols
needed to determinize nondeterministic input-driven pushdown automata (NIDPDA)
working over a fixed alphabet
is determined precisely.
It is proved that in the worst case exactly $2^{n^2}$ states are needed
to determinize an $n$-state NIDPDA,
and the proof uses witness automata
with a stack alphabet $\Gamma = \{0,1\}$ working on strings over a $4$-symbol input alphabet
(Only an asymptotic lower bound was known before in the case of a fixed alphabet).
Also, the impact
of NIDPDA determinization on the size of stack alphabet
is determined precisely for the first time:
it is proved
that $s(2^{n^2}-1)$ stack symbols are necessary in the worst case to determinize
an $n$-state NIDPDA working over an input alphabet of size $s+5$ with $s$ left brackets
(The previous lower bound was only asymptotic in the number of states and did not depend on the number of left brackets).
\end{abstract}

\section{Introduction}

Input-driven pushdown automata, also known as visibly pushdown automata, 
are a model of computation equipped with a stack. An automaton has finitely many states. 
It reads an input string from the left to the right changing its state. The automaton
also uses an infinite memory in the form of stack with restricted access.
An input alphabet is $\Sigma = \Sigma_0 \cup \Sigma_{+1} \cup \Sigma_{-1}$,
with $\Sigma_0$ containing neutral symbols,
$\Sigma_{+1}$ left brackets and $\Sigma_{-1}$ right brackets.
When an automaton reads a left bracket, it pushes onto the stack
a symbol of its stack alphabet; when it sees a right bracket, it looks at the top symbol of the stack
and pops it out; and when the automaton processes a neutral symbol, it makes a transition
without looking at the stack.

Deterministic input-driven pushdown automata (DIDPDA)
were invented by Mehlhorn~\cite{Mehlhorn}.
He also proved that every language defined by an $n$-state DIDPDA can be recognized
by an algorithm that uses $O(\frac{(\log n)^2}{\log \log n})$ bits of memory
and works in polynomial time. 
Von Braunm\"uhl and Verbeek~\cite{vonBraunmuehl_Verbeek} 
considered a nondeterministic version of input-driven pushdown automata (NIDPDA)
that at every step may have several possible actions. Such an automaton accepts a string
if there is at least one accepting computation on it.
A natural question immediately follows: whether deterministic and nondeterministic input-driven pushdown automata are equal in power.
Von Braunm\"uhl and Verbeek~\cite{vonBraunmuehl_Verbeek} made the first determinization construction: for an $n$-state NIDPDA, 
working over an alphabet $\Sigma = \Sigma_0 \cup \Sigma_{+1} \cup \Sigma_{-1}$,
they constructed a DIDPDA with $2^{n^2}$ states
and with $2^{n^2}|\Sigma_{+1}|$ stack symbols recognizing the same language.
Also von Braunm\"uhl and Verbeek~\cite{vonBraunmuehl_Verbeek} improved
the result by Mehlhorn~\cite{Mehlhorn}: they have shown that
any language defined by an $n$-state DIDPDA is recognized by an algorithm
working in logarithmic memory. Later Rytter~\cite{Rytter} created a simpler algorithm
for this task also using $\log n$ bits of memory.

Alur and Madhusudan~\cite{AlurMadhusudan,AlurMadhusudan2009} 
reintroduced the model of input-driven pushdown automata
under the name of \emph{visibly pushdown automata}
and obtained many important results on these automata.
They proved that the class of languages recognized by these automata
is closed under intersection, union, concatenation and the Kleene star.
They also defined input-driven pushdown automata
that work on infinite strings and investigated properties of this variant of the model.
Alur and Madhusudan~\cite{AlurMadhusudan,AlurMadhusudan2009} established
the first lower bound on the number of states needed for NIDPDA determinization:
for each $n$, they constructed an $n$-state NIDPDA with the fixed input alphabet such that
any deterministic automaton recognizing the same language has at least $2^{\Omega(n^2)}$ states;
this lower bound is asymptotically tight.
Furthermore, Alur and Madhusudan~\cite{AlurMadhusudan,AlurMadhusudan2009}
studied decidability and complexity of input-driven pushdown automata
and showed that universality and inclusion problems
for NIDPDA are NEXP-complete.
There is some current research on decision problems for NIDPDA. 
Han, Ko and Salomaa~\cite{HanKoSalomaa} built an algorithm that for a given NIDPDA
decides in polynomial time whether its path size is finite, that is, whether
a number of leaves in a tree of its computations on every string is bounded
by a common constant. They also proved that deciding whether the
path size of a given NIDPDA is less than a given number is EXP-complete.

The research on NIDPDA determinization was continued by
Okhotin, Piao and Salomaa~\cite{OkhotinPiaoSalomaa}, who bounded from below
not only the number of states in the deterministic automaton
but also the number of stack symbols it uses.
They showed that using $2^{\Omega(n^2)}$ stack symbols
can be necessary to determinize an $n$-state NIDPDA.
Furthermore, Okhotin, Piao and Salomaa~\cite{OkhotinPiaoSalomaa}
and later Okhotin and Salomaa~\cite{OkhotinSalomaa_operations}
studied the state complexity of different operations on DIDPDA.
Jir\'askov\'a and Okhotin~\cite{JiraskovaOkhotin_idpda_sc} continued improving bounds
for operations on DIDPDA, and in addition proved
that in the worst case one needs $2^{n^2}$ states to determinize an $n$-state NIDPDA.
This lower bound is precise,
however, witness automata used by Jir\'askov\'a and Okhotin~\cite{JiraskovaOkhotin_idpda_sc}
work over an input alphabet of size exponential in $n$.
So the problem of determining the exact state complexity of 
NIDPDA determinization in the case of a bounded alphabet has remained open.

The state complexity of determinization was investigated for variants of the classical model
of input-driven pushdown automata.
Nguyen Van Tang and Ogawa~\cite{VanTang_Ogawa} introduced event-clock input-driven
pushdown automata and proved that these automata can be determinized.
Later Ogawa and Okhotin~\cite{OgawaOkhotin} defined a direct determinization construction
for these automata, establishing an upper bound: it is enough to use
$2^{n^2}$ states, $2^{n^2+k}|\Sigma_{+1}|$ stack symbols and $k$ clock constraints
to determinize an $n$-state nondeterministic event-clock input-driven pushdown automaton 
with $k$ clock constraints.
Furthermore, Ogawa and Okhotin~\cite{OgawaOkhotin} proved a lower bound
to this transformation which is asymptotically precise both in the number of states 
and in the size of a stack alphabet. Rose and Okhotin~\cite{RoseOkhotin} first
considered probabilistic input-driven pushdown automata
and determined asymptotically precisely the state complexity of
determinization for this model.
Kutrib, Malcher and Wendlandt~\cite{KutribMalcherWendlandt_transducer_driven}
defined a variant of input-driven pushdown automata
in which every input string is read twice: first,
a deterministic sequential transducer determines the type of each symbol (in this model
an alphabet is not initially split into left brackets, right brackets and neutral symbols),
and then the string is read by an input-driven pushdown automaton.
Kutrib et al.~\cite{KutribMalcherWendlandt_transducer_driven} proved
that such input-driven pushdown automata with transducers are stronger than without transducers,
determinized such automata and established their closure properties.

In this paper I improve the bounds on the complexity of determinization
for classical input-driven pushdown automata (NIDPDA)
both in the number of states
and in the size of the stack alphabet.
In Section~\ref{section_determinization_fixed_alph_states_2_n_2},
it is proved that in the worst case one can need $2^{n^2}$ states
to determinize an $n$-state NIDPDA that uses $2$ stack symbols
and works over a $4$-symbol input alphabet. This is the first precise
lower bound on the number of states needed for NIDPDA determinization
with bounded alphabet.

In Section~\ref{section_one_opening_bracket}, the precise lower bound
$|\Sigma_{+1}|(2^{n^2}-1)$ on the number of stack symbols
in a deterministic automaton that recognizes a language defined by an $n$-state NIDPDA
is proved in a special case of only one left bracket ($\Sigma_{+1} = \{{<}\}$).
The witness nondeterministic automata in this proof have stack alphabets growing linearly in $n$,
and the input alphabet is bounded.

Finally, in Section~\ref{section_lower_bound_many_opening_brackets}
I establish the exact lower bound on the complexity of NIDPDA determinization,
for any number of left brackets in the input alphabet fewer than $2^{n^2}$,
both in the number of states and in the number of stack symbols:
in the worst case one needs $2^{n^2}$ states and $|\Sigma_{+1}|(2^{n^2}-1)$
stack symbols to determinize an $n$-state NIDPDA.
Moreover, these examples of NIDPDA that are hard to determinize work over input alphabets
that do not depend on $n$ and use stack alphabets growing linearly in $n$
and logarithmically in the number of left brackets.
This lower bound is precise,
since the known upper bound of $|\Sigma_{+1}|\cdot 2^{n^2}$
stack symbols is improved to $|\Sigma_{+1}|(2^{n^2}-1)$ stack symbols in Section~\ref{section_definitions}.

\section{Input-driven pushdown automata}\label{section_definitions}

Input-driven pushdown automata, also known as visibly pushdown automata,
were investigated a lot.
These automata were first invented by Mehlhorn~\cite{Mehlhorn}; 
later von Braunm\"uhl and Verbeek~\cite{vonBraunmuehl_Verbeek}
introduced the nondeterministic version of input-driven pushdown automata.
 
This paper uses the definition given by Alur and Madhusudan~\cite{AlurMadhusudan}
but with one difference: Alur and Madhusudan~\cite{AlurMadhusudan}
allow computations on ill-nested strings,
whereas in the definition used in this paper, input strings must be well-nested,
as in the first definitions of these automata.

A nondeterministic automaton will be defined first, and a deterministic one
is its special case.

\begin{definition}[Mehlhorn~\cite{Mehlhorn}, von Braunm\"uhl and Verbeek~\cite{vonBraunmuehl_Verbeek},
Alur and Madhusudan~\cite{AlurMadhusudan}]
\emph{A nondeterministic input-driven pushdown automaton (NIDPDA)}
is a sextuple $A = (\Sigma, Q, \Gamma, Q_0, (\delta_a)_{a \in \Sigma}, F)$, where
\begin{itemize}
\item
	$\Sigma = \Sigma_0 \cup \Sigma_{+1} \cup \Sigma_{-1}$ is a finite input alphabet
	split into three disjoint sets: $\Sigma_0$ contains neutral symbols, 
	$\Sigma_{+1}$ consists of left brackets and
	$\Sigma_{-1}$ has right brackets;
\item
	$Q$ is a finite set of states of the automaton;
\item
	$\Gamma$ is a finite stack alphabet;
\item
	$Q_0 \subseteq Q$ is a subset of initial states;
\item
	$(\delta_a)_{a \in \Sigma}$ are functions that for each symbol of the alphabet
	define possible actions of the automaton at this symbol:
	\begin{itemize}
	\item
		for a neutral symbol $a \in \Sigma_0$, the function $\delta_a \colon Q \to 2^Q$
		for each state gives a set of possible next states of the automaton;
	\item
		for a left bracket $a \in \Sigma_{+1}$, the function $\delta_a \colon Q \to 2^{Q \times \Gamma}$
		for each state assumed at a symbol $a$ specifies the set of pairs $(q,s)$
		in which the automaton
		can make a transition forward in the state $q$ pushing the symbol $s$ onto the stack;
	\item
		for a right bracket $a \in \Sigma_{-1}$, 
		the function $\delta_a \colon Q \times \Gamma \to 2^Q$ 
		for a state and for a symbol popped out of the stack
		gives a set of all possible next states;
	\end{itemize}
\item
	$F \subseteq Q$ is a subset of accepting states.
\end{itemize}

Inputs of the automaton $A$ are well-nested strings over an alphabet $\Sigma$.
A string $w$ is called \emph{well-nested} if it has as many left brackets as right brackets 
and if every prefix contains at least as many left brackets as right brackets.

Let $w = a_1\ldots a_{\ell}$ be a well-nested string.
A \emph{computation} of the automaton $A$ on the string $w$
is a sequence $(p_0,\alpha_0)$, \ldots, $(p_{\ell}, \alpha_{\ell})$ 
of pairs of a state and of stack contents,
with the neighbouring pairs related to each other as follows.

The automaton $A$ begins reading the string in one of the initial states with an empty stack:
the initial pair $(p_0,\alpha_0)$ must have the form $(q_0,\epsilon)$, where $q_0 \in Q_0$
is an initial state.
Let the automaton have processed the first $i$ symbols and let a pair $(p_i,\alpha_i)$
describe its state and the stack at the moment.
Then, the next pair $(p_{i+1}, \alpha_{i+1})$ should be reachable from the previous
pair by the transition by the next symbol of the string. Consider three cases.
\begin{itemize}
\item
	If the next symbol is neutral: $a_{i+1} = c$, for $c \in \Sigma_0$,
	then the stack remains the same: $\alpha_{i+1} = \alpha_i$,
	and the next state is obtained from the previous
	state by a transition: $p_{i+1} \in \delta_c(p_i)$.
\item
	If the next symbol in the string is a left bracket: $a_{i+1} = {<}$, 
	for ${<} \in \Sigma_{+1}$,
	then a new stack symbol is pushed onto the stack at the current step
	and the next pair must have the form $(p_{i+1},\alpha_{i+1}) = (r,\alpha_is)$,
	for a state $r \in Q$ and for a stack symbol $s \in \Gamma$ such that $(r,s) \in \delta_{<}(p_i)$.
\item
	If the next symbol is a right bracket: $a_{i+1} = {>}$,
	for ${>} \in \Sigma_{-1}$, then the symbol at the top of the stack is popped;
	the stack is not empty at the current moment, since the string $w$ is well-nested.
	Let $s \in \Gamma$ be a symbol at the top of the stack, that is, let $\alpha_i = \beta s$.
	Then, the automaton removes the top stack symbol: $\alpha_{i+1} = \beta$, 
	and changes its state to one of the possible next states, while seing the popped symbol: 
	$p_{i+1} \in \delta_{>}(p_i,s)$.
\end{itemize}

The automaton finishes its computation on a string in a state $p_{\ell}$ with an empty stack $\alpha_{\ell}$ (the stack is empty since the string $w$ is well-nested).
If the final state is accepting: $p_{\ell} \in F$, then
the computation $(p_0,\alpha_0)$, \ldots, $(p_{\ell}, \alpha_{\ell})$ of the automaton $A_n$
on the string $w$ is called \emph{accepting}.
The string $w$ is said to be accepted by the automaton
if there is at least one accepting computaion on this string.
And the automaton $A$ defines the language $L(A)$ consisting of 
all well-nested strings over the alphabet $\Sigma$ accepted by the automaton.
\end{definition}

An input-driven pushdown automaton is called \emph{deterministic} (DIDPDA),
if it has a unique initial state, $|Q_0| = 1$,
and if in every situation it has exactly one possible action:
\begin{itemize}
\item
	$|\delta_a(q)| = 1$, for all $q \in Q$ and $a \in \Sigma_0 \cup \Sigma_{+1}$;
\item
	$|\delta_a(q,s)| = 1$, for all $q \in Q$, $s \in \Gamma$ and $a \in \Sigma_{-1}$.
\end{itemize}
Simplified notation is used in the deterministic case:
it is said that a deterministic automaton 
has a unique initial state $q_0$ and all functions give not a set of possible actions but
a deterministically defined action: $\delta_a \colon Q \to Q$, for $a \in \Sigma_0$, 
$\delta_a \colon Q \to Q \times \Gamma$,
for $a \in \Sigma_{+1}$, and $\delta_a \colon Q \times \Gamma \to Q$, for $a \in \Sigma_{-1}$.

A DIDPDA is complete, that is, all functions $(\delta_a)_{a \in \Sigma}$
are fully defined. There are some definitions of DIDPDA in the literature that allow
rejecting in the middle of the string. 
The bounds proved for complete and for partial variants of the definition
usually differ by one state.

It is known that NIDPDA can be determinized:
von Braunm\"uhl and Verbeek~\cite{vonBraunmuehl_Verbeek} for an arbitrary $n$-state NIDPDA
constructed a deterministic automaton recognizing the same language 
with $2^{n^2}$ states and with $2^{n^2}|\Sigma_{+1}|$
stack symbols.

The number of stack symbols in the determinization construction
can be slightly improved to $(2^{n^2}-1)|\Sigma_{+1}|$.
This small improvement was not needed before,
but since I prove exact lower bounds in Sections~\ref{section_one_opening_bracket} and~\ref{section_lower_bound_many_opening_brackets},
it is useful to make the upper bound match these lower bounds.

\begin{theorem}\label{theorem_determinization_stack_alph_slightly_better}
Let $A$ be an $n$-state NIDPDA, 
working over an alphabet $\Sigma = \Sigma_0 \cup \Sigma_{+1} \cup \Sigma_{-1}$.
Then, there is a DIDPDA with $2^{n^2}$ states and with
$|\Sigma_{+1}|(2^{n^2}-1)$ stack symbols that recognizes the language $L(A)$.
\end{theorem}
\begin{proof}
Consider the modern presentation of the determinization construction
by von Braunm\"uhl and Verbeek~\cite{vonBraunmuehl_Verbeek} given in the survey by
Okhotin and Salomaa~\cite[Thm.~1]{OkhotinSalomaa}.
Let $B$ be a deterministic automaton obtained from $A$, as in this survey.
The idea of the construction is that the deterministic automaton $B$ calculates
the behaviour relation $R \subseteq Q \times Q$ of the nondeterministic automaton
on a current well-nested segment of a string, and when the automaton $B$
sees a left bracket, it pushes this relation onto the stack along with the bracket.
This is how $2^{n^2}$ states and $2^{n^2}|\Sigma_{+1}|$ stack symbols occur.

On the other hand, if the automaton $B$ enters the state $\emptyset$ in any situation,
then the behavior relation of a current well-nested segment is empty,
and this segment cannot be passed by the nondeterministic automaton $A$.
So, the state $\emptyset$ means that the string should be rejected by $B$.
Then, this state is made a rejecting state---the transition function is modified
so that all transitions from the state $\emptyset$ lead again to this state $\emptyset$.
Is it useful to push onto the stack the empty relation $R = \emptyset$?
If the automaton $B$ pushes a symbol of the form $(\emptyset, {<})$,
then a string contains a well-nested segment that cannot be passed through by $A$
and the string is rejected. Then, the transition that pushes $(\emptyset, {<})$ onto the stack
can be modified. The automaton can instead assume the rejecting state $\emptyset$ 
and push onto the stack a symbol $(R, {<})$ with any non-empty relation $R$,
and the automaton will reject the string as well.
After the transitions of the automaton $B$ are changed this way,
the stack symbols of the form $(\emptyset, {<})$ are not pushed onto the stack anymore.
These symbols can be removed from the stack alphabet, leaving
$(2^{n^2}-1)|\Sigma_{+1}|$ stack symbols.
\end{proof}

\section{The exact bound on the number of states}
\label{section_determinization_fixed_alph_states_2_n_2}

Alur and Madhusudan~\cite{AlurMadhusudan,AlurMadhusudan2009} 
established the first lower bound on the state complexity
of the determinization of NIDPDA.
They proved that some nondeterministic input-driven pushdown automata with $n$ states
require at least $2^{\Omega(n^2)}$ states in equivalent deterministic automata.
Alur and Madhusudan~\cite{AlurMadhusudan,AlurMadhusudan2009} used NIDPDA examples
with a fixed input alphabet and with stack alphabet growing linearly in $n$.
Later, Jir\'askov\'a and Okhotin~\cite{JiraskovaOkhotin_idpda_sc} 
obtained the exact lower bound of $2^{n^2}$ states, using
nondeterministic automata with input alphabet exponential in $n$.

In this section, the exact lower bound of $2^{n^2}$ states on the state complexity 
of determinization
is proved for the first time in the case of a fixed input alphabet.
Moreover, the stack alphabet of nondeterministic automata
is also fixed and is of size $2$.
This is the minimal size of the stack alphabet
that allows an automaton to get any information from the stack.

\begin{theorem}\label{theorem_states_precise_lb_alph2}
For each $n \geqslant 1$, there is an $n$-state NIDPDA 
$A_n = (\Sigma, Q, \Gamma, q_0, (\delta_a)_{a \in \Sigma}, F)$,
over a $4$-symbol input alphabet
$\Sigma_{+1} = \{{<}\}$, $\Sigma_{-1} = \{{>}\}$, $\Sigma_0 = \{{-},\#\}$,
with one initial state,
and with stack alphabet $\Gamma = \{0,1\}$,
such that any DIDPDA recognizing the language $L(A_n)$
has at least $2^{n^2}$ states.
\end{theorem}
\begin{proof}
Let $n$ be fixed. The desired automaton $A_n$ is defined as follows.
The set of states is $Q = \{0,\ldots, n-1\}$,
with the initial state $0$. All states are accepting: $Q = F$.
The input alphabet $\Sigma$ and the stack alphabet $\Gamma$
have already been defined in the theorem statement.
And the symbols of the input alphabet act in the following way.

The symbol `$\#$' contains all the nondeterminism of the automaton $A_n$,
the automaton can move from every state to every state by this symbol:
\begin{align*}
	\delta_{\#}(i) &= \{0,\ldots, n-1\}, && \text{for } i \in \{0, \ldots, n-1\}.
\intertext{%
The automaton works deterministically on all other symbols of $\Sigma$.
The symbol `${-}$' decreases the state of $A_n$ by $1$ modulo $n$:
}
	\delta_-(i) &= \{(i-1) \bmod n\}, && \text{for } i \in \{0, \ldots, n-1\}.
\intertext{%
By the left bracket `${<}$', the automaton does not change the state 
and pushes the information, whether the current state is $0$ or not, onto the stack, 
that is, for a state $i$, it pushes $\mathop{\mathrm{sgn}} i$, which is $0$ if the state is $0$, 
and is $1$ otherwise:
}
	 \delta_{<}(i) &= \{(i,\mathop{\mathrm{sgn}} i)\}, && \text{for } i \in \{0, \ldots, n-1\}.
\intertext{%
By the right bracket `${>}$', the automaton either stays in the same state or rejects.
If the current state is not $0$, or if the symbol in the stack is not $0$,
the automaton $A_n$ continues in the same state.
And if the current state and the symbol at the top of the stack both equal $0$,
then the automaton rejects:
}
	\delta_{>}(i,s) &= \{i\}, 
		&& \text{for } i \in \{0,\ldots, n-1\}, \; s \in \{0, 1\},\; (i,s) \neq (0,0).
\end{align*}

Now the automaton $A_n$ has been defined.

If $n=1$, the theorem is trivial, because a deterministic automaton needs
at least $2^1 = 2$ states: an accepting state to accept the empty string and a rejecting state
to reject the string ${<}{>}$.
Then, let $n$ be at least $2$.

It should be proved that a deterministic automaton
must use at least $2^{n^2}$ states to recognize the language $L(A_n)$.

\begin{figure}[t]
	\centerline{\includegraphics[scale=1.1]{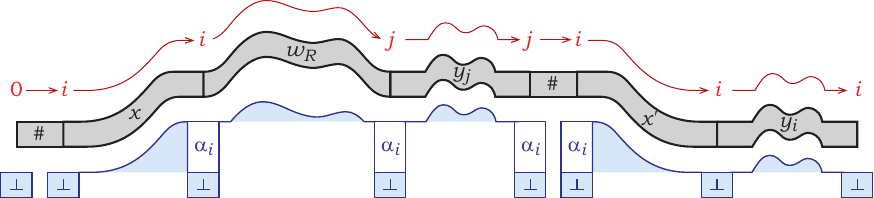}}
	\caption{The accepting computation of the automaton $A_n$ on the string $\#xw_Ry_j\#x'y_i$,
	where $(i,j) \in R$.}
	\label{f:idpda_det_stack2_string}
\end{figure}

First, strings with specific behaviour of the automaton $A_n$ are constructed.

\begin{claim}\label{claim_w_R_y_i_x_x'}
For each relation $R \subseteq Q \times Q$, there is a well-nested string $w_R \in \Sigma^*$,
such that for all $i,j \in Q$, the automaton $A_n$ can start reading the string $w_R$
in the state $i$ and finish reading it in the state $j$ if and only if $(i,j) \in R$.

For each state $i \in Q$, there is such a well-nested string $y_i \in \Sigma^*$,
that if the automaton $A_n$ enters $y_i$ in some state other than $i$, then it rejects,
and if $A_n$ enters the string $y_i$ in the state $i$,
then it can leave the string in the state $i$ and cannot leave the string $y_i$ in any other state.

There are two strings $x \in \{{-},{<}\}^*$ and $x' \in \{{-},{>}\}^*$, such that
$x$ has as many left brackets as there are right brackets in $x'$.
If the automaton $A_n$ begins reading $x$ in the state $i \in Q$,
then it leaves this string in the same state $i$ and pushes some string $\alpha_i$ onto the stack.
And if $A_n$ enters the string $x'$ in a state $i' \in Q$
having the string $\alpha_i$ on the top of the stack, then, if $i' \neq i$, it rejects,
and if $i' = i$, it leaves the string $x'$ in the state $i$.
\end{claim}

The plan of the proof of Theorem~\ref{theorem_states_precise_lb_alph2}
is to build strings with special properties defined in Claim~\ref{claim_w_R_y_i_x_x'}
using encoding over a small aphabet, 
and then to consider computations of the automaton $A_n$ 
on strings of the form $\#xw_Ry_j\#x'y_i$, for $R \subseteq Q\times Q$ and $i, j \in Q$.
It will be proved, that if $(i,j) \in R$, the automaton $A_n$ accepts the string $\#xw_Ry_j\#x'y_i$,
as in Figure~\ref{f:idpda_det_stack2_string}. And if $(i,j) \notin R$, then
there will be no accepting computations of $A_n$ on this string.
As in the earlier lower bound proofs~\cite{AlurMadhusudan2009},
the final step of the plan is to show that any deterministic automaton
should remember the relation $R$ in its state after reading the string $w_R$
to be ready to give the correct answer for every possible pair $(i,j) \in Q \times Q$
in the suffix $y_j\#x'y_i$.

The conditions on strings $w_R$ in Claim~\ref{claim_w_R_y_i_x_x'}
can be reformulated as follows.

For a well-nested string $w \in \Sigma^*$, one can define a \emph{behaviour relation} 
$R(w) \subseteq Q \times Q$: the pair $(i,j)$ is in $R(w)$ if and only if
there is a computation of the nondeterministic automaton $A_n$ on the string $w$,
that begins in the state $i$ and ends by leaving the string in the state $j$.
In these terms, for each relation $R \subseteq Q \times Q$, one wants to construct
a well-nested string $w_R$ with the behaviour relation $R(w_R) = R$.

For example, for a full behaviour relation, one can take a string
${\#}$ as $w_{Q \times Q}$, because $R({\#}) = Q \times Q$.
To construct strings with all possible relations, it is enough to learn
how to eliminate an arbitrary pair of states from the relation of a string.
This is done in the following claim.

\begin{claim}\label{claim_R_u_iwv_j}
There are strings $u_i \in \{{-},{<}\}^*$, for all $i \in Q$,
and $v_j \in \{{-},{>}\}^*$, for all $j \in Q$,
such that for each well-nested string $w \in \Sigma^*$,
and for all states $i,j \in Q$,
the equality $R(u_iwv_j) = R(w)\setminus\{(i,j)\}$ holds.

Furthermore, strings $u_i$ and $v_j$ satisfy the following conditions. 
\begin{itemize}
\item
	Each string $u_i$, for $i \in Q$, has exactly one left bracket,
	and there is exactly one right bracket in each string $v_j$, for $j \in Q$.
\item
	If the automaton $A_n$ begins reading the string $u_i$, for $i \in Q$, or $v_j$, for $j \in Q$,
	in some state, then it can leave the string only in this state.
	On the string $v_j$ the automaton can reject in some cases;
	it never rejects while reading the string $u_i$.
\end{itemize}
\end{claim}

Strings $u_i$ and $v_j$ are defined as follows:
\begin{align*}
u_i &= ({-})^i{<}({-})^{n-i}, && \text{ for } i \in Q;\\
v_j &= ({-})^j{>}({-})^{n-j}, && \text{ for } j \in Q.
\end{align*}
There are no symbols `${\#}$' in these strings, and therefore
the automaton $A_n$ works deterministically on them.
When the automaton $A_n$ enters a string $u_i$ in a state $i'$,
first, it passes $i$ symbols `${-}$' decreasing its state $i$ times,
and comes to the symbol `${<}$' in a state $(i'-i) \bmod n$.
If $i' = i$, this state is $0$ and the automaton pushes $0$ onto the stack
at the left bracket `${<}$', otherwise it pushes $1$ onto the stack.
Finally, it reads the last $n-i$ symbols `${-}$'
and finishes reading the string $u_i$ in the state $i'$.

Then, consider computations of the automaton $A_n$ on a string $v_j$. Let $A_n$
enter this string in a state $j' \in Q$ with a symbol $s \in \{0,1\}$ on the top of the stack.
After reading the first $j$ symbols `${-}$' the automaton comes to the symbol `${>}$'
in a state $(j'-j) \bmod n$.
If $s = 0$ and $j' = j$ at the same time, then $A_n$ rejects,
and if the symbol on the top of the stack is not $0$
or the state of the automaton at the symbol `${>}$' is not $0$,
the automaton safely passes through the right bracket.
And after reading the remaining $n-j$ symbols `${-}$' the automaton recovers the state $j'$.

How the behaviour relation on a well-nested string $w$
will change if one wraps it in a pair of strings $u_i$ and $v_j$, for $i,j \in Q$,
that is, if the string $w$ is replaced with $u_iwv_j$?

Can the pair $(i,j)$ lie in the relation $R(u_iwv_j)$? Consider
any computation of $A_n$ on the string $u_iwv_j$ that enters the string in the state $i$
and leaves the string in the state $j$.
The automaton begins reading $u_i$ in the state $i$, and so it pushes $0$ at the top of the stack.
Since the substring $w$ is well-nested, the automaton will pop this symbol $0$ out of the stack 
while reading the substring $v_j$, and because the substring $v_j$
does not change the state of the automaton, the automaton $A_n$ enters 
the substring $v_j$ in the state $j$. Then, it rejects and $(i,j) \notin R(u_iwv_j)$.
The relation $R(u_iwv_j)$ cannot contain any pairs that are not in $R(w)$,
because the strings $u_i$ and $v_j$ do not change the state of the automaton,
and therefore the automaton should enter the substring $w$ in the state $i'$ 
and leave this substring in the state $j'$ to enter the whole string $u_iwv_j$ 
in the state $i'$ and leave it in the state $j'$.
Each pair $(i',j') \in R(w)$, that is not equal to $(i,j)$, lies in the relation $R(u_iwv_j)$,
because there is the following computation on the string $u_iwv_j$. First, the automaton
enters the substring $u_i$ in the state $i'$ and leaves it in the state $i'$,
while pushing $1$ onto the stack if $i' \neq i$, and $0$ otherwise.
Then, the automaton reads the substring $w$ and changes its state from $i'$ to $j'$,
it can do this, since $(i',j') \in R(w)$. Finally, it begins reading the string $v_j$ in the state $j'$.
Since $(i',j') \neq (i,j)$, at this moment the automaton cannot at the same time be in the state $j$
and have $0$ at the top of the stack,
so the automaton does not reject while reading $v_j$
and finishes its computation by leaving the string $u_iwv_j$ in the state $j'$.
Therefore, $R(u_iwv_j) = R(w) \setminus \{(i,j)\}$.

Claim~\ref{claim_R_u_iwv_j} is proved.

Now it is time to construct all special strings in Claim~\ref{claim_w_R_y_i_x_x'}.
Strings $w_R$ for all $R \subseteq Q \times Q$ can be obtained from 
a string $\#$ with a full behaviour relation, using Claim~\ref{claim_R_u_iwv_j}.
Let $R \subseteq Q \times Q$ be an arbitrary relation. Let 
$(i_1,j_1)$ ,\ldots, $(i_k,j_k)$ be all pairs of states that are not in $R$.
Then, the string $w_R$ can be defined as:
\begin{equation*}
w_R = u_{i_k}u_{i_{k-1}}\ldots u_{i_1}{\#}v_{j_1}\ldots v_{j_{k-1}}v_{j_k}.  
\end{equation*}
By Claim~\ref{claim_R_u_iwv_j}, the behaviour relation on this string is 
obtained from $Q \times Q$ by eliminating pairs $(i_1,j_1)$ ,\ldots, $(i_k,j_k)$,
so it equals $R$. Also, this string $w_R$ is well-nested, 
since each string $u_{i_t}$, for $t = 1, \ldots, k$, has exactly one left bracket,
and each string $v_{j_t}$, for $t = 1, \ldots, k$,
has exactly one right bracket.

\begin{figure}[t]
	\centerline{\includegraphics[scale=1.1]{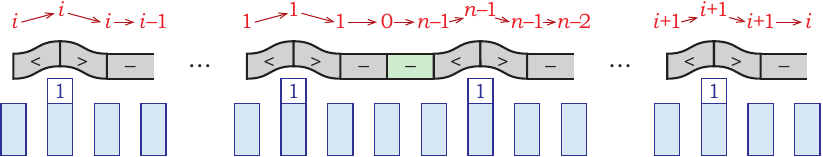}}
	\caption{The automaton $A_n$ begins reading the string $y_i$ in the state $i$
	and deterministically leaves this string in the same state $i$.}
	\label{f:idpda_det_stack2_yi}
\end{figure}

Next, the well-nested string $y_i \in \Sigma^*$, for $i \in Q$, can be constructed as:
\begin{equation*}
y_i = w_R, \text{ where } R = \{(i,i)\}.
\end{equation*}
There is another way to define the string $y_i$, explicitly and without symbols `${\#}$':
\begin{equation*}
y_i = ({<}{>}{-})^i{-}({<}{>}{-})^{n-i},
\end{equation*}
In this construction, each pair of brackets ${<}{>}$ forbids the current state to be
equal to $0$ and as a result it is prohibited to enter the string in any state other than $i$.
Figure~\ref{f:idpda_det_stack2_yi} shows how the automaton $A_n$
reads the string $y_i$, starting in the state $i$. If the automaton enters the string $y_i$
in some state $i' \neq i$, then it will be in the state $0$ after reading the first $i'$ symbols `${-}$'.
And since $i' \neq i$, there is a substring ${<}{>}$ in the string $y_i$ 
after the $i'$-th symbol `${-}$';
the automaton reads this substring in the state $0$ and rejects.

It remains to construct the strings $x \in \{{-},{<}\}^*$ and $x' \in \{{-},{>}\}^*$.
Consider the diagonal relation $R = \{(0,0), (1,1),\ldots, (n-1,n-1)\}$.
The string with this relation is 
$w_R = u_{i_k}u_{i_{k-1}}\ldots u_{i_1}{\#}v_{j_1}\ldots v_{j_{k-1}}v_{j_k}$,
where $(i_1,j_1),\ldots,(i_k,j_k)$ are all pairs of states that are not in $R$.
Then, the strings $x$ and $x'$ are defined as:
\begin{align*}
x &= u_{i_k}u_{i_{k-1}}\ldots u_{i_1},\\
x' &= v_{j_1}\ldots v_{j_{k-1}}v_{j_k}.
\end{align*}
Then the string $x$ is defined over an alphabet $\{{-},{<}\}$ and contains exactly
$k$ left brackets, whereas $x'$ is a string over an alphabet $\{{-},{>}\}$ 
and has exactly $k$ right brackets.
The strings $x$ and $x'$ have no symbols `${\#}$', so the automaton $A_n$
works on these strings deterministically. For each state $i \in Q$,
there is a uniquely defined string $\alpha_i \in \Gamma^*$, which is pushed onto the stack
when the automaton reads the string $x$ from the state $i$.
Since the strings $u_i$, for $i \in Q$, preserve the state of the automaton $A_n$,
the string $x$ also cannot change the state of the automaton.
Analogously, the automaton cannot change its state by reading $x'$,
but it may reject on $x'$, because the strings $v_j$, for $j \in Q$,
have these properties.

It shall be proved that if the automaton reads the string $x'$ with the string $\alpha_i$
at the top of the stack, then it can move through this string in the state $i$, but the automaton will
reject the string if it enters $x'$ in any other state.
Consider the behaviour relation of the automaton $A_n$ on the string $x{\#}x'$.
The symbol `${\#}$' allows the automaton to change each state to each state,
whereas the strings $x$ and $x'$ preserve the state of the automaton. 
Thus, the pair $(i,j) \in Q \times Q$ is in the relation $R(x{\#}x')$ if and only if 
the automaton $A_n$ first enters the string $x$ in the state $i$ and pushes the string $\alpha_i$
onto the stack, and then enters the string $x'$ in the state $j$ having the string $\alpha_i$
on the top of the stack and does not reject.
On the other hand, by the definition, the relation $R(x{\#}x')$ is the diagonal relation:
$R(x{\#}x') = \{(0,0), (1,1),\ldots, (n-1,n-1)\}$.
So, for each $i \in Q$, since $(i,i) \in R(x{\#}x')$, 
the automaton $A_n$ can enter the string $x'$ in the state $i$
with the string $\alpha_i$ at the top of the stack, and finish reading $x'$ in the state $i$.
Also, the automaton should have no computations on the string $x'$ 
with $\alpha_i$ at the top of the stack 
that begin and end in some state $j \neq i$, since $(i,j) \notin R(x{\#}x')$, for $j \neq i$.
Therefore, the constructed strings $x$ and $x'$ are as desired
and this finishes the proof of Claim~\ref{claim_w_R_y_i_x_x'}.

Now all special strings have been constructed, and it remains to prove
that the nondeterministic automaton $A_n$ accepts a string $\#xw_Ry_j\#x'y_i$
if and only if $(i,j) \in R$, and that every deterministic automaton needs many states
to do the same.

\begin{claim}\label{claim_A_n_on_string_i_j_in_R}
Let $i, j \in Q$ and $R \subseteq Q \times Q$, 
let the well-nested strings $w_R, y_i, y_j \in \Sigma^*$
and the strings $x \in \{{-},{<}\}^*$, $x' \in \{{-},{>}\}^*$
be constructed as in Claim~\ref{claim_w_R_y_i_x_x'}.
Then, the automaton $A_n$ accepts the string $\#xw_Ry_j\#x'y_i$ if and only if $(i,j) \in R$.
\end{claim}

Let $(i,j)$ be in $R$. Then there is the following accepting computation
of the automaton $A_n$ on the string $\#xw_Ry_j\#x'y_i$, 
illustrated in Figure~\ref{f:idpda_det_stack2_string}. 
At the first symbol `${\#}$' the automaton guesses the state $i$.
It enters the substring $x$ in the state $i$ and pushes the string $\alpha_i$ onto the stack
while reading $x$. Then, it leaves the substring $x$ in the same state $i$ and enters
the substring $w_R$.
Since $(i,j) \in R$, the automaton can leave the substring $w_R$ in the state $j$
after entering it in the state $i$. So, the automaton changes the state from $i$ to $j$ 
and finishes reading the substring $w_R$ in the state $j$. The next substring $y_j$
prohibits all states except $j$, but the automaton is in the state $j$,
and it safely passes this substring. Then, at the other symbol `$\#$'
the automaton $A_n$ nondeterministically chooses the state $i$.
Since the substring $w_Ry_j{\#}$ is well-nested, the string $\alpha_i$
that has been pushed onto the stack while reading the substring $x$,
is at the top of the stack at this moment.
The automaton moves through the substring $x'$ popping the string $\alpha_i$ out of the stack,
and comes to the next substring in the same state $i$. The stack is empty after reading $x'$.
And finally the automaton $A_n$ reads the substring $y_i$ in the state $i$
and leaves the whole string $\#xw_Ry_j\#x'y_i$ in the state $i$.
Therefore, the automaton accepts the string, because all states are accepting.

Now let the automaton $A_n$ accept a string $\#xw_Ry_j\#x'y_i$,
for some $i,j \in Q$ and $R \in Q \times Q$. It shall be proved that $(i,j) \in R$.
Consider an arbitrary accepting computation of the automaton $A_n$ 
on the string $\#xw_Ry_j\#x'y_i$.
Let $i'$ be the state guessed by the automaton at the first symbol `${\#}$',
let $j'$ be the state assumed by $A_n$ after reading the substring $w_R$,
and let the state $i''$
be chosen at the symbol `${\#}$' before the substring $x'y_i$.
The automaton always gets out of the substring $x'$ in the state in which it enters this substring,
and also the automaton rejects if it enters the substring $y_i$ in any state other than $i$.
Therefore, $i'' = i$. Not to reject on the substring $x'$, the automaton should enter it
in the same state in which it has entered the substring $x$,
so $i'' = i' = i$.
The automaton will reject while reading the substring $y_j$ if it starts reading it in a state
other than $j$. Thus, $j' = j$.
However, since the substring $x$ cannot change the state of the automaton,
the automaton $A_n$ enters the substring $w_R$ in the state $i$
and leaves it in the state $j$. This means that $(i,j) \in R$.

It has been shown that the nondeterministic automaton $A_n$ works correctly on strings
of the form $\#xw_Ry_j\#x'y_i$, and accepts such a string if and only if $(i,j) \in R$.
Now it will be proved that no deterministic automaton recognizes the language $L(A_n)$
using fewer than $2^{n^2}$ states.

\begin{claim} \label{claim_Didpda_2_n2}
Let $A$ be a deterministic input-driven pushdown automaton that recognizes the language $L(A_n)$. 
Then the automaton $A$ has at least $2^{n^2}$ states. 
\end{claim}

By the definition, DIDPDA is complete, so the automaton $A$ cannot reject in the middle of
the string.
The automaton $A$ accepts a string of the form $\#xw_Ry_j\#x'y_i$ if and only if $(i,j) \in R$,
since it recognizes the same language as $A_n$, and since the automaton $A_n$ has this property
by Claim~\ref{claim_A_n_on_string_i_j_in_R}. Here all the strings $w_R$, for $R \subseteq Q \times Q$, $y_i$, for $i \in Q$, $x$ and $x'$ are constructed by Claim~\ref{claim_w_R_y_i_x_x'}. 
Let $q_R$, for any relation $R \subseteq Q \times Q$, be the state assumed by the
deterministic automaton $A$ after reading the substring $\#xw_R$ from the initial state.
It will be shown that all states $q_R$, for $R \in Q \times Q$, are pairwise distinct,
then the automaton $A$ will have at least $2^{n^2}$ such states.

For the sake of a contradiction, let $R_1, R_2 \subseteq Q \times Q$ be two different
relations with $q_{R_1} = q_{R_2}$. Since $R_1 \neq R_2$, 
there is a pair of states $(i,j) \in Q \times Q$
that lies in exactly one of the relations $R_1$, $R_2$.
Then, the automaton $A$ accepts only one of the strings $\#xw_{R_1}y_j\#x'y_i$ and
$\#xw_{R_2}y_j\#x'y_i$. 
While reading the prefix ${\#}x$ the deterministic automaton $A$
always pushes onto the stack the same string $\alpha$.
And since $q_{R_1} = q_{R_2}$ and since the substrings $w_{R_1}$ and $w_{R_2}$ are well-nested,
the automaton $A$ gets out of substrings $\#xw_{R_1}$ and $\#xw_{R_2}$
in the same state $q_{R_1} = q_{R_2}$ with the stack $\alpha$.
Then, in its computations on the strings $\#xw_{R_1}y_j\#x'y_i$ and
$\#xw_{R_2}y_j\#x'y_i$ the deterministic automaton enters the suffix $y_j\#x'y_i$
in the state $q_{R_1} = q_{R_2}$ with the stack $\alpha$.
Thus, the automaton either accepts both strings or rejects both,
this is a contradiction.

Therefore, all the states $q_R$, for $R \subseteq Q \times Q$,
are pairwise distinct and the deterministic automaton $A$
has at least $2^{n^2}$ states.
Theorem~\ref{theorem_states_precise_lb_alph2} is proven.
\end{proof}

Note that if an $n$-state nondeterministic input-driven pushdown automaton
uses a one-symbol stack alphabet, then it can be determinized like a simple DFA
by the well-known subset construction using only $2^n$ states.
Therefore, using two stack symbols in a stack alphabet $\Gamma$ in the automata in
Theorem~\ref{theorem_states_precise_lb_alph2} is optimal.

\section{The lower bound on the number of stack symbols: one left bracket} \label{section_one_opening_bracket}

It has been proven in Section~\ref{section_determinization_fixed_alph_states_2_n_2}
that, in the case of a fixed alphabet, $2^{n^2}$ states can be necessary to
determinize an $n$-state nondeterministic automaton.

However, the size of an input-driven pushdown automaton is described not only 
by the number of its states but also by the number of stack symbols it uses.
The determinization construction by von Braunm\"uhl and Verbeek~\cite{vonBraunmuehl_Verbeek},
for an $n$-state NIDPDA, gives a deterministic automaton with $2^{n^2}$ states
and $2^{n^2}|\Sigma_{+1}|$ stack symbols.
This upper bound has been slightly improved to $(2^{n^2}-1)|\Sigma_{+1}|$ stack symbols
and the same $2^{n^2}$ states 
in Theorem~\ref{theorem_determinization_stack_alph_slightly_better} 
in Section~\ref{section_definitions}.

How optimal is this determinization construction in the number of stack symbols?
The best lower bound of $2^{\frac{n^2}{c}}$ stack symbols, with a constant $c > 1$, 
was obtained by Okhotin, Piao and Salomaa~\cite{OkhotinPiaoSalomaa}.
They used witness automata working on strings over a bounded input alphabet
and with a stack alphabet of size linear in $n$.

The goal is to prove an exact lower bound on the number of stack symbols
needed for NIDPDA determinization,
and preferably using a linearly growing stack alphabet
and working over a fixed input alphabet.
To do this, one needs to work efficiently with small alphabets
to avoid division by a constant $c$,
and also needs to learn how to use several left brackets
to multiply the bound by $|\Sigma_{+1}|$.

In this section, the exact bound is proved in a simpler case of only one left bracket,
and in the next Section~\ref{section_lower_bound_many_opening_brackets}
it will be shown how to use any number of left brackets less than $2^{n^2}$
to multiply the number of stack symbols in a deterministic automaton by $|\Sigma_{+1}|$.

\begin{theorem}\label{theorem_stack_alph_precise_lb_1_opening_bracket}
For each $n \geqslant 1$, there is an $n$-state NIDPDA 
$B_n = (\Sigma, Q, \Gamma, q_0, (\delta_a)_{a \in \Sigma}, F)$,
with one initial state, with a stack alphabet of size $|\Gamma| = 2n+2$, working over
a $5$-symbol input alphabet: 
$\Sigma_{+1} = \{{<}\}$, $\Sigma_{-1} = \{{>},{\gg}\}$, $\Sigma_0 = \{{-},\#\}$,
such that any DIDPDA recognizing a language $L(B_n)$
has at least $2^{n^2}$ states and at least $2^{n^2}-1$ stack symbols.
\end{theorem}

\begin{proof}
Consider the automaton $A_n$
defined in the proof of Theorem~\ref{theorem_states_precise_lb_alph2}.
It is an $n$-state automaton with $Q = \{0,1,\ldots, n-1\}$; the state $0$
is initial and all states are accepting;
the automaton has stack alphabet $\{0,1\}$ and works over an input alphabet 
$\Sigma \setminus \{{\gg}\}$.
The automaton $B_n$ will be obtained from $A_n$ by adding a new right bracket `${\gg}$',
new stack symbols $\widehat{i}$ and $\overrightarrow{i}$, for $i \in Q$,
and some new transitions.

The stack alphabet of the automaton $B_n$ is
\begin{equation*}
\Gamma = \{0,1\} \cup \set{\widehat{i}}{i \in Q} \cup \set{\overrightarrow{i}}{i \in Q}.
\end{equation*}

Transitions by symbols `${-}$', `$\#$', `${>}$' in the automaton $B_n$ remain
the same as in $A_n$.
The automaton $A_n$ does not change its state on a left bracket `${<}$',
and pushes $0$ onto the stack if the state is $0$, and $1$ if the state is not $0$
(in other words, $A_n$ pushes $\mathop{\mathrm{sgn}} i$ in a state $i$).
The automaton $B_n$ can either do the same, or alternatively decide to change its state
(maybe, to the same state) and to push onto the stack
either the old state $i$ as $\widehat{i}$,
or the new state $j$ as $\overrightarrow{j}$.
\begin{align*}
	\delta_{<}(i) &= \{(i,\mathop{\mathrm{sgn}} i)\} \cup \set{(j,\widehat{i})}{j \in Q}
	\cup \set{(j,\overrightarrow{j})}{j \in Q}, 
		&& \text{for } i \in \{0,\ldots, n-1\}
\end{align*}
If a new stack symbol is popped out of the stack at an old right bracket `${>}$',
then the automaton rejects.
If $0$ or $1$ is at the top of the stack when the automaton $B_n$ reads a double bracket `${\gg}$',
then it also rejects.
If a symbol $\widehat{i}$, for $i \in Q$,
is popped out of the stack at a double bracket `${\gg}$', then, if the current state is $0$, 
the automaton moves to the state $i$, and otherwise rejects.
\begin{align*}
	\delta_{\gg}(0,\widehat{i}) &= \{i\}, 
		&& \text{for } i \in Q
\intertext{%
And if a stack symbol is of the form $\overrightarrow{j}$, for $j \in Q$,
then at a double bracket `${\gg}$' the automaton changes the state $1$ to the state $j$
and rejects in all states other than $1$.
}
	\delta_{\gg}(1,\overrightarrow{j}) &= \{j\}, 
		&& \text{for } j \in Q
\end{align*}

Now the automaton $B_n$ has been defined.

If $B_n$ begins reading some well-nested string without double brackets `$\gg$'
in some state $i \in Q$ and leaves the string in some state $j \in Q$,
then in this computation the automaton $B_n$ makes a transition of $A_n$ 
at each left bracket `${<}$'.
Indeed, if $B_n$ chooses to push onto the stack any new stack symbol 
while reading the left bracket `${<}$', then it will later pop this symbol out of the stack
at the matching old right bracket `${>}$' and reject.
Therefore, the next property holds.

\begin{claim}\label{claim_B_n_works_as_A_n_on_well_nested_strings_without_new_brackets}
For each well-nested string without double brackets `${\gg}$'
and for all states $i,j \in Q$,
the automaton $B_n$ can change its state from $i$ to $j$ by reading this string
if and only if so can $A_n$.
\end{claim}

Using Claim~\ref{claim_B_n_works_as_A_n_on_well_nested_strings_without_new_brackets},
one can easily prove that determinization of $B_n$
requires at least as many states as determinization of $A_n$.

\begin{claim}\label{claim_B_n_2_pow_n_pow_2_states}
Every DIDPDA recognizing the language $L(B_n)$ has at least $2^{n^2}$ states.
\end{claim}

First, by Claim~\ref{claim_B_n_works_as_A_n_on_well_nested_strings_without_new_brackets},
automata $A_n$ and $B_n$ accept the same sets of strings without double brackets `${\gg}$',
that is, $L(B_n) \cap (\Sigma \setminus \{{\gg}\})^* = L(A_n)$.
It is also used here that $A_n$ and $B_n$ have the same initial and accepting states.
For the sake of a contradiction, assume that there is a deterministic automaton $A$
that recognizes the language $L(B_n)$ using less than $2^{n^2}$ states.
Then, one can delete the double bracket `$\gg$' from the alphabet of $A$
getting a new deterministic automaton $A'$, recognizing the language $L(A_n)$
with fewer than $2^{n^2}$ states. 
And it is impossible by Theorem~\ref{theorem_states_precise_lb_alph2}.

If $n=1$, then Claim~\ref{claim_B_n_2_pow_n_pow_2_states} is enough
to prove Theorem~\ref{theorem_stack_alph_precise_lb_1_opening_bracket},
since the stack alphabet of any DIDPDA
has at least $2^{1^2}-1 = 1$ symbol.
From this moment, let $n$ be at least $2$;
the following proof uses the existence of states $0$ and $1$.

Let $y_i$, for $i \in Q$, and $w_R$, for $R \subseteq Q \times Q$,
be well-nested strings constructed in Claim~\ref{claim_w_R_y_i_x_x'}
in the proof of Theorem~\ref{theorem_states_precise_lb_alph2}.
These strings contain no double brackets `${\gg}$',
therefore the automaton $B_n$ works on them as $A_n$.
And the next claim is a direct corollary of Claim~\ref{claim_w_R_y_i_x_x'}.

\begin{claim}\label{claim_B_n_on_y_i_and_on_wR}
All strings $y_i$, for $i \in Q$, and $w_R$, for $R \subseteq Q \times Q$,
are well-nested, and the automaton $B_n$ works on them as follows.
\begin{enumerate}
\item
	If $B_n$ enters a string $y_i$, for $i \in Q$, in any state other than $i$,
	then it rejects, and if it enters this string in the state $i$,
	then it can leave it in the state $i$ and cannot leave it in any other state.
\item
	The automaton $B_n$ can enter a string $w_R$ in a state $i$
	and leave it in a state $j$ if and only if $(i,j) \in R$.
\end{enumerate}
\end{claim}

It shall be proved that every deterministic automaton
recognizing the language of the automaton $B_n$
has at least $2^{n^2}-1$ stack symbols.
To this end, special strings comprised of any substrings $w_R$, with $R \subseteq Q \times Q$, 
and $y_i$, with $i \in Q$, and also of symbols `${<}$', `${\gg}$', `$\#$',
are constructed.
For a sequence of $m$ non-empty relations $R_1, R_2, \ldots, R_m \subseteq Q \times Q$,
a string $f_{R_1,\ldots, R_m}$ is defined:
\begin{equation*}
f_{R_1,\ldots, R_m} = {<}w_{R_1}{<}w_{R_2}\ldots{<}w_{R_m}{<}.
\end{equation*}
Also, for all $i,j \in Q$, for each $m \geqslant 1$ and for each $k = 1, \ldots, m$,
a string $g_{i,j,k,m}$ is defined:
\begin{equation*}
g_{i,j,k,m} = (\#{\gg})^{m-k}\#y_0{\gg}y_j\#y_1{\gg}y_i(\#{\gg})^{k-1}
\end{equation*}
The goal is to prove that the nondeterministic automaton $B_n$
accepts a string $f_{R_1,\ldots, R_m}g_{i,j,k,m}$
if and only if $(i,j) \in R_k$,
and that any deterministic automaton needs at least $2^{n^2}-1$ stack symbols to do the same.

\begin{claim}\label{claim_B_n_on_special_string}
Let $m \geqslant 1$ and $k = 1, \ldots, m$ be integers, 
let $R_1, R_2, \ldots, R_m \subseteq Q \times Q$
be non-empty relations, let $i,j \in Q$ be two states.
Then, the automaton $B_n$ accepts a string $f_{R_1,\ldots,R_m}g_{i,j,k,m}$
if and only if $(i,j) \in R_k$.
\end{claim}

\begin{figure}[t]
	\centerline{\includegraphics[scale=1.1]{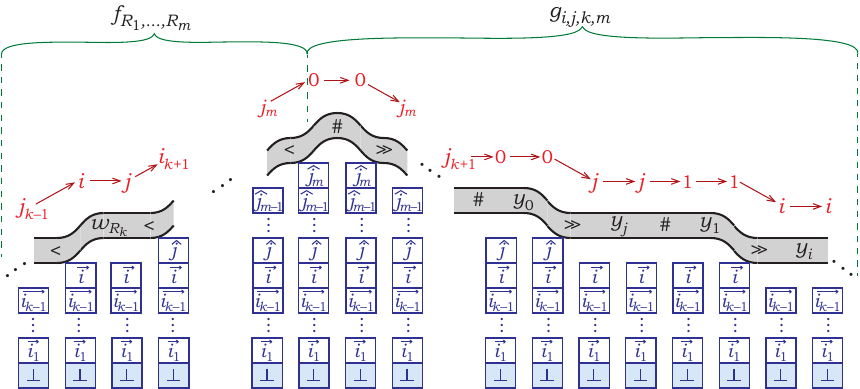}}
	\caption{An accepting computation of the automaton $B_n$ 
	on a string $f_{R_1,\ldots,R_m}g_{i,j,k,m}$, for $(i,j) \in R_k$. 
	First, it is shown in the figure how the automaton moves through
	the substring ${<}w_{R_k}{<}$ of  
	$f_{R_1,\ldots,R_m} = {<}w_{R_1}{<}w_{R_2}\ldots{<}w_{R_m}{<}$,
	pushing onto the stack $\protect\overrightarrow{i}\widehat{j}$.
	Then $B_n$ enters the second part $g_{i,j,k,m}$ of the string,
	that begins in the figure with a symbol `${\#}$' at the top,
	and it is illustrated how $B_n$ gets through a substring 
	$\#y_0{\gg}y_j\#y_1{\gg}y_i$ of $g_{i,j,k,m}$
	popping symbols $\widehat{j}$ and $\protect\overrightarrow{i}$ out of the stack.}
	\label{f:idpda_det_1left_f_g}
\end{figure}

First, let $(i,j)$ lie in $R_k$. Then there is the following accepting computation of the automaton $B_n$ on a string $f_{R_1,\ldots,R_m}g_{i,j,k,m}$ (see Figure~\ref{f:idpda_det_1left_f_g}).
Since relations $R_1$, \ldots, $R_m$ are non-empty, one can choose pairs $(i_1,j_1) \in R_1$,
$(i_2,j_2) \in R_2$, \ldots, $(i_m, j_m) \in R_m$, taking $i_k = i$ and $j_k = j$.
For convenience, let also $j_0 = 0$ and $i_{m+1} = 0$.
The string $f_{R_1,\ldots,R_m} = {<}w_{R_1}{<}w_{R_2}\ldots{<}w_{R_m}{<}$
has $m+1$ unmatched left brackets and the automaton $B_n$
starts its computation at the first of them in the state $0$.
The computation will be constructed so that the automaton will get to each $t$-th
unmatched left bracket in the state $j_{t-1}$, for all $t = 1, \ldots, m+1$.
On $t$-th unmatched bracket the automaton $B_n$ moves from the state $j_{t-1}$ to the state $i_t$,
pushing onto the stack either the symbol $\overrightarrow{i_t}$, for $t \leqslant k$,
or the symbol $\widehat{j_{t-1}}$, for $t > k$. 
Next, if $t \neq m+1$,
the automaton enters the substring $w_{R_t}$ in the state $i_t$.
This is a well-nested string, so reading it does not change the stack.
Since $(i_t,j_t) \in R_t$, there is a computation on $w_{R_t}$
that begins in the state $i_t$ and ends in the state $j_t$.
This computation is used as a part of an accepting computation 
on the full string.
As a result, the automaton gets from the state $j_{t-1}$ at the $t$-th unmatched left bracket
to the state $j_t$ at the next unmatched left bracket.
Thus, the correct computation of the automaton $B_n$ on the substring $f_{R_1,\ldots,R_m}$
has been constructed, and it ends with the automaton leaving the substring in the state $0$
with the following stack contents:
\begin{equation*}
\overrightarrow{i_1}\overrightarrow{i_2} \ldots \overrightarrow{i_k} 
\widehat{j_k} \widehat{j_{k+1}} \ldots \widehat{j_m}.
\end{equation*}

Next, the automaton $B_n$ reads a string 
$g_{i,j,k,m} = (\#{\gg})^{m-k}\#y_0{\gg}y_j\#y_1{\gg}y_i(\#{\gg})^{k-1}$,
and the computation should be continued to make the automaton accept the full string.
The automaton can pass by a substring $(\#{\gg})^{m-k}$ assuming
the state $0$ after reading each symbol `$\#$'.
Since the last $m-k$ symbols pushed onto the stack
are $ \widehat{j_{k+1}}$, \ldots $\widehat{j_m}$, 
which are symbols of the form $\widehat{q}$, for $q \in Q$,
the automaton can pop these symbols out of the stack in the state $0$ at double right brackets `${\gg}$'
without rejecting.
Then, $B_n$ needs to get through a substring $\#y_0{\gg}y_j\#y_1{\gg}y_i$.
This can be done as follows: at the first symbol `${\#}$' the automaton
assumes the state $0$, in this state it moves through $y_0$ without rejecting
and enters the first double bracket `${\gg}$' of this substring.
The automaton pops the symbol $\widehat{j_k}$ out of the stack in the state $0$
at the double bracket `${\gg}$' and moves to the state $j_k = j$.
This allows the automaton $B_n$ to pass through the substring $y_j$.
At the second symbol `$\#$' the automaton chooses the state $1$
and passes $y_1$. Then it pops the symbol $\overrightarrow{i_k}$ out of the stack
in the state $1$ at `${\gg}$' and changes its state to $i_k = i$.
In this state $i$ it safely moves through $y_i$.
It remains to read the last substring $(\#{\gg})^{k-1}$, and this can be done
by choosing the state $1$ at every symbol `$\#$'. The remaining string in the stack
is $\overrightarrow{i_1}\overrightarrow{i_2} \ldots \overrightarrow{i_{k-1}}$,
and the automaton can pop symbols of this form at double brackets `${\gg}$' in the state $1$.
Thus, the automaton $B_n$ completes its computation on the string $f_{R_1,\ldots,R_m}g_{i,j,k,m}$
and accepts, since all states are accepting.

Now let $(i,j)$ be not in $R_k$. It shall be proved, that the string $f_{R_1,\ldots,R_m}g_{i,j,k,m}$
is not accepted by $B_n$. Consider an arbitrary computation of the automaton $B_n$
on the string $f_{R_1,\ldots,R_m}g_{i,j,k,m}$. There are exactly $m+1$ unmatched
left brackets in a substring $f_{R_1,\ldots,R_m}$, and if the automaton did not reject
while reading this substring, then it will leave the substring
with some string $\alpha = s_1\ldots s_{m+1}$ of length $m+1$ in the stack.
Symbols $s_k$ and $s_{k+1}$ were pushed onto the stack
at the first and at the last left brackets in the substring ${<}w_{R_k}{<}$.
If $s_k = \overrightarrow{i}$ and $s_{k+1} = \widehat{j}$, 
then the automaton $B_n$ leaves the first symbol `${<}$'
and enters the substring $w_{R_k}$ in the state $i$
and leaves this substring and enters the last left bracket in the state $j$.
Since $(i,j) \notin R_k$, it is impossible to enter $w_{R_k}$ in the state $i$
and leave it in the state $j$. Therefore, either $s_k \neq \overrightarrow{i}$,
or $s_{k+1} \neq \widehat{j}$.
It will be shown that in both cases the automaton rejects while reading the substring
$g_{i,j,k,m} = (\#{\gg})^{m-k}\#y_0{\gg}y_j\#y_1{\gg}y_i(\#{\gg})^{k-1}$.
The symbol $s_{k+1}$ is popped out of the stack
while reading the substring $\#y_0{\gg}y_j$. 
Strings $y_0$ and $y_j$ wrapping the double right bracket `${\gg}$'
require that the automaton enters `${\gg}$' in the state $0$ and leaves it in the state $j$.
Such a change of a state at `${\gg}$' is possible only
with the symbol $\widehat{j}$ at the top of the stack. Therefore, if $s_{k+1} \neq \widehat{j}$,
the automaton rejects. 
The second case is when $s_k \neq \overrightarrow{i}$. 
The symbol $s_k$ is popped out of the stack while reading the substring $\#y_1{\gg}y_i$,
here the double right bracket `${\gg}$' is wrapped in substrings $y_1$ and $y_i$, 
and not to reject the automaton $B_n$ should change
the state $1$ to the state $i$ at the double bracket `${\gg}$'.
But this is impossible since the symbol at the top of the stack is not $\overrightarrow{i}$.
Therefore, if $(i,j) \notin R_k$,
the automaton $B_n$ rejects the string $f_{R_1,\ldots,R_m}g_{i,j,k,m}$ in any computation.

Claim~\ref{claim_B_n_on_special_string} has been proven.
It remains to show that any deterministic automaton needs a lot of stack symbols
to work as $B_n$.

\begin{claim}\label{claim_A_on_special_string_needs_many_stack_alph_symbols}
Let DIDPDA $A$ work over an alphabet $\Sigma$
and let it accept a string $f_{R_1,\ldots,R_m}g_{i,j,k,m}$ if and only if $(i,j) \in R_k$, 
for all $m \geqslant 1$,
for all non-empty relations $R_1, R_2, \ldots, R_m \subseteq Q \times Q$,
for all states $i,j \in Q$ and for all $k \in \{1, \ldots, m\}$.
Then, there are at least $2^{n^2}-1$ stack symbols in the stack alphabet of the automaton $A$.
\end{claim}

Let $N$ be the number of states in the automaton $A$ 
and let $M$ be the size of its stack alphabet.
Then the automaton $A$ finishes reading each string of the form $f_{R_1,\ldots,R_m}$
in one of $N$ states with a string of length $m+1$ in the stack,
since a string $f_{R_1,\ldots,R_m}$ has exactly $m+1$ unmatched left brackets.
For a fixed $m$, there are at most $N \cdot M^{m+1}$
possible outcomes of reading a string $f_{R_1,\ldots,R_m}$,
for $m$ arbitrary non-empty relations.
On the other hand, for a fixed $m$, a number of strings of the form $f_{R_1,\ldots,R_m}$
is $(2^{n^2}-1)^m$, since each of $m$ non-empty relations 
$R_1$, \ldots, $R_m \subseteq Q \times Q$
can be chosen in $2^{n^2}-1$ ways.

For the sake of a contradiction, assume that the deterministic automaton $A$
has a small number of stack symbols: $M < 2^{n^2}-1$.
Since the number of states $N$ of the automaton $A$ is fixed
and $M < 2^{n^2}-1$, one can choose $m$ large enough,
so that the number of possible outcomes of reading a substring $f_{R_1,\ldots,R_m}$
will be fewer than the number of such substrings: $N \cdot M^{m+1} < (2^{n^2}-1)^m$.
Therefore, there are two distinct sequences of non-empty relations $R_1$, \ldots, $R_m$
and $S_1$, \ldots, $S_m$, such that the automaton $A$ finishes reading the strings
$f_{R_1,\ldots,R_m}$ and $f_{S_1,\ldots,S_m}$ in the same state $q$
and with the same stack contents $\alpha$.

Since the sequences of relations $R_1$, \ldots, $R_m$ and $S_1$, \ldots, $S_m$ 
are different, there is a pair of states $(i,j) \in Q \times Q$
and there is a number $k \in \{1,\ldots, m\}$, such that the pair $(i,j)$ lies in
exactly one of the relations $R_k$ and $S_k$.
Without loss of generality, one can assume that $(i,j) \in R_k$ and $(i,j) \notin S_k$.
Then the deterministic automaton $A$ accepts the string $f_{R_1,\ldots,R_m}g_{i,j,k,m}$
and rejects the string $f_{S_1,\ldots,S_m}g_{i,j,k,m}$.
However, on both strings the automaton enters the second part $g_{i,j,k,m}$
in the state $q$ with the string $\alpha$ in the stack,
and therefore it finishes reading both strings $f_{R_1,\ldots,R_m}g_{i,j,k,m}$
and $f_{S_1,\ldots,S_m}g_{i,j,k,m}$ in the same state,
and either accepts both strings,
ot rejects both. This is a contradiction,
and Claim~\ref{claim_A_on_special_string_needs_many_stack_alph_symbols} is proved.

By Claim~\ref{claim_B_n_on_special_string},
the nondeterministic automaton $B_n$ accepts a string $f_{R_1,\ldots,R_m}g_{i,j,k,m}$
if and only if $(i,j) \in R_k$. Then, every deterministic automaton
recognizing the language of $B_n$ does the same and satisfies the conditions
of Claim~\ref{claim_A_on_special_string_needs_many_stack_alph_symbols}.
Therefore it has at least $2^{n^2}-1$ stack symbols, and
Theorem~\ref{theorem_stack_alph_precise_lb_1_opening_bracket} has been proven.
\end{proof}

\section{The lower bound on the number of stack symbols: several left brackets}
\label{section_lower_bound_many_opening_brackets}

The number of stack symbols sufficient for determinizing
an $n$-state NIDPDA is $(2^{n^2}-1)|\Sigma_{+1}|$,
established in Section~\ref{section_definitions}.
In the case of only one left bracket,
the exact lower bound on the size of stack alphabet
needed for determinization is $2^{n^2}-1$,
proved in Section~\ref{section_one_opening_bracket}.
How to multiply the number of stack symbols needed for determinization
by the number of left brackets, if $|\Sigma_{+1}| > 1$? 
The exact lower bound in the case of several left brackets is proved
in the next theorem.   

\begin{theorem}\label{theorem_stack_alph_precise_lb_several_opening_brackets}
For any integers $n$ and $s$, such that $n \geqslant 1$ and $1 \leqslant s \leqslant 2^{n^2}$,
there is an $n$-state NIDPDA 
$B_{n,s} = (\Sigma, Q, \Gamma, q_0, (\delta_a)_{a \in \Sigma}, F)$,
with one initial state, with a stack alphabet of the size $|\Gamma| = 2+2n+\lfloor\log_2 (2s-1)\rfloor$
and working over an $(s+5)$-symbol input alphabet: 
$\Sigma_{+1} = \{{<}_0, {<}_1,\ldots,{<}_{s-1}\}$, 
$\Sigma_{-1} = \{{>},{\gg},{\ggg}\}$, 
$\Sigma_0 = \{{-},\#\}$,
such that every DIDPDA recognizing the language $L(B_{n,s})$
has at least $2^{n^2}$ states and at least $s(2^{n^2}-1)$ stack symbols.
\end{theorem}
\begin{proof}
Let $n \geqslant 2$ (a degenerate case of $n=1$ will be handled at the end of the proof).
Also let $s \geqslant 2$ (if $s = 1$, then one can take the automaton $B_n$ from
Theorem~\ref{theorem_stack_alph_precise_lb_1_opening_bracket} as $B_{n,s}$).
Then $\lfloor\log_2 (2s-1)\rfloor = \lfloor\log_2 (s-1)\rfloor+1$.

The automaton $B_{n,s}$ is constructed as a more complicated version of the automaton $B_n$
in Theorem~\ref{theorem_stack_alph_precise_lb_1_opening_bracket}.
It has a set of states $Q = \{0,\ldots, n-1\}$,
with the initial state $0$ and with all states accepting, as the automaton $B_n$.
The input alphabet of $B_{n,s}$ is defined in the statement 
of Theorem~\ref{theorem_stack_alph_precise_lb_several_opening_brackets}.
The stack alphabet is
\begin{equation*}
\Gamma = \{0,1\} \cup \set{\widehat{i}}{i \in Q} \cup \set{\overrightarrow{i}}{i \in Q} \cup \set{\text{\textcircled{$x$}}}{x = 0,\ldots,\lfloor\log_2 (s-1)\rfloor}.
\end{equation*}
Stack symbols from the set 
$\{0,1\} \cup \set{\widehat{i}}{i \in Q} \cup \set{\overrightarrow{i}}{i \in Q}$
are already used by the automaton $B_n$ in Theorem~\ref{theorem_stack_alph_precise_lb_1_opening_bracket};
the automaton $B_{n,s}$ has also some new stack symbols: $\text{\textcircled{$x$}}$, for $x \in \{0,\ldots,\lfloor\log_2 (s-1)\rfloor\}$.
And $|\Gamma| = 2+2n+\lfloor\log_2 (s-1)\rfloor+1 = 2+2n+\lfloor\log_2 (2s-1)\rfloor$,
as required.

Next, the transition function of the automaton $B_{n,s}$ is defined.

The possible transitions of $B_{n,s}$
at the neutral symbols `$\#$' and `$-$' and at the old right brackets `${>}$' and `${\gg}$'
coincide with the transitions of $B_n$ at these symbols.

For each left bracket `${<}_\ell$', for $\ell \in \{0, \ldots, s-1\}$,
the automaton $B_{n,s}$ has all transitions that are defined for $B_n$ at the left bracket `${<}$'.
Also there are new transitions,
that arbitrarily change the state of the automaton and push onto the stack
the position of any unary bit in the binary representation of the number $\ell$ of the left bracket.
Denote the coefficient at $2^x$ in the binary representation of the number $\ell$
as $\ell[x]$, and denote the transition function of $B_n$ as $\delta^{(B_n)}$.
Then the transition function of the automaton $B_{n,s}$ at left brackets
is defined as:
\begin{align*}
	\delta_{{<}_\ell}(q) &= \delta^{(B_n)}_{<}(q) \cup \set{(r,\text{\textcircled{$x$}})}{r \in Q, \; \ell[x] = 1},
		&&\text{for } \ell \in \{0,\ldots, s-1\}, \; q \in Q. 
\end{align*}

It remains to define the transition function of $B_{n,s}$
at a new triple right bracket `${\ggg}$'.
No transitions are defined at this bracket
for old stack symbols from sets $\{0,1\}$, 
$\set{\widehat{i}}{i \in Q}$ and $\set{\overrightarrow{i}}{i \in Q}$.
If a new stack symbol of the form $\text{\textcircled{$x$}} \in \{0,\ldots,\lfloor\log_2 (s-1)\rfloor\}$
is at the top of the stack, then the automaton $B_{n,s}$ at a triple bracket `$\ggg$'
can only change its state from the state $(x \bmod n)$ to the state $\lfloor\frac{x}{n}\rfloor$.
Such a definition of transitions at a triple bracket `${\ggg}$'
allows the automaton to extract all the information about a symbol $\text{\textcircled{$x$}}$
from the stack.
The automaton cannot simply put $x$ in its state while reading a triple bracket `${\ggg}$',
since, for $s = 2^{n^2}$, the number $x$ can reach the value of $\lfloor\log_2 (2^{n^2}-1)\rfloor=n^2-1$, and the automaton has only $n$ states. Thus, a number $x$ at the top of the stack 
is read in two parts: 
the state $(x \bmod n)$ should be assumed before reading the bracket `$\ggg$',
and the state $\lfloor\frac{x}{n}\rfloor$ emerges after reading this bracket.
Since $0 \leqslant x \leqslant n^2-1$, it holds that $x = \lfloor\frac{x}{n}\rfloor\cdot n+(x \bmod n)$.
\begin{align*}
	\delta_{{\ggg}}(x\bmod  n,\; \text{\textcircled{$x$}}) &= \{\Big\lfloor\frac{x}{n}\Big\rfloor\}
\end{align*}

The automaton $B_{n,s}$ has been completely defined.

If the automaton $B_{n,s}$ pops out of the stack a new stack symbol 
of the form $\text{\textcircled{$x$}}$ at one of the old right brackets `${>}$' and `${\gg}$',
it immediately rejects. So, new stack symbols can be popped only at a
triple right bracket `${\ggg}$'.
Thus, accepting computations of $B_{n,s}$ on well-nested strings
without triple brackets `${\ggg}$' are the same as accepting computations
of $B_n$, if each left bracket in the input alphabet of $B_{n,s}$
is considered as the only left bracket `${<}$' of the automaton $B_n$.
Then one can bound from below the number of states needed for determinization of
the automaton $B_{n,s}$.
\begin{claim}\label{claim_B_n_s_many_states}
Each DIDPDA recognizing the language $L(B_{n,s})$
has at least $2^{n^2}$ states.
\end{claim}

Indeed, if one eliminates left brackets ${<_1}$, \ldots, ${<_s}$
from the alphabet of the DIDPDA,
leaving only the symbol `${<_0}$' as `${<}$',
and if one also deletes the triple right bracket `${\ggg}$',
then the resulting deterministic automaton will recognize exactly
the language of the automaton $B_n$ and will have the same number of states
as the original deterministic automaton. And the language $L(B_n)$
cannot be recognized by a DIDPDA with fewer than $2^{n^2}$ states.

Furthermore, since the automaton $B_{n,s}$ works on well-nested strings without triple brackets `$\ggg$',
as the automaton $B_n$, if all left brackets of $B_{n,s}$ are considered 
as the bracket `${<}$' of $B_n$, one can define strings $w_R$, for $R \subseteq Q \times Q$,
and $y_i$, for $i \in Q$, for the automaton $B_{n,s}$.
These strings are constructed like in Claim~\ref{claim_w_R_y_i_x_x'}
in the proof of Theorem~\ref{theorem_states_precise_lb_alph2};
the left bracket `${<}_0$' is used as the symbol `${<}$'.
These strings are well-nested and do not contain double and triple right brackets `${\gg}$' and `${\ggg}$',
therefore the automaton $B_{n,s}$ works on them as $B_n$,
and the automaton $B_n$ works on them as the automaton $A_n$ 
in Theorem~\ref{theorem_states_precise_lb_alph2}.
Then, computations of $B_{n,s}$ on these strings are described by the next claim
that directly follows from Claims~\ref{claim_w_R_y_i_x_x'}
and~\ref{claim_B_n_on_y_i_and_on_wR}.

\begin{claim}\label{claim_B_n_s_on_y_i_and_on_wR}
The automaton $B_{n,s}$ works on well-nested strings $y_i$, for $i \in Q$,
and $w_R$, for $R \subseteq Q \times Q$, as follows.
\begin{enumerate}
\item
	If the automaton enters a string $y_i$, for $i \in Q$, in the state $i$,
	then it can leave this string in the state $i$ and cannot leave it in any other state.
	If the automaton $B_{n,s}$ begins reading $y_i$ in any state other than $i$,
	it rejects.
\item
	For every two states $i,j \in Q$,
	the automaton $B_{n,s}$ can begin reading a string $w_R$ in the state $i$
	and leave this string in the state $j$, if and only if $(i,j) \in R$.
\end{enumerate}
\end{claim}

Strings $f_{R_1,\ldots, R_m}$ and $g_{i,j,k,m}$ were defined for the automaton $B_n$
in Theorem~\ref{theorem_stack_alph_precise_lb_1_opening_bracket}.
Each string $f_{R_1,\ldots, R_m}$
has exactly $m+1$ unmatched left brackets and $m$ intermediate substrings
with non-empty behaviour relations $R_1$, \ldots, $R_m$.
Now, more elaborate strings will be defined,
in which every unmatched left bracket 
can be any numbered bracket: ${<_0}$, ${<_1}$, \ldots, ${<_{s-1}}$.
Such a string is determined not only by a sequence of $m$ non-empty relations,
but also by a sequence of $m+1$ left bracket numbers.
For every integer $m \geqslant 1$, 
for a sequence of $m$ non-empty relations $R_1, R_2, \ldots, R_m \subseteq Q \times Q$
and for a sequence of indices $\ell_1, \ldots, \ell_{m+1} \in \{0, \ldots, s-1\}$,
the following string is defined:
\begin{equation*}
f_{R_1,\ldots, R_m,\ell_1,\ldots,\ell_{m+1}} = {<_{\ell_1}}w_{R_1}{<_{\ell_2}}w_{R_2}\ldots{<_{\ell_m}}w_{R_m}{<_{\ell_{m+1}}}.
\end{equation*}
Strings $g_{i,j,k,m}$, for $i,j \in Q$, for $m \geqslant 1$ and for $k = 1, \ldots, m$,
are exactly the same as for the automaton $B_n$:
\begin{equation*}
g_{i,j,k,m} = (\#{\gg})^{m-k}\#y_0{\gg}y_j\#y_1{\gg}y_i(\#{\gg})^{k-1}
\end{equation*}

If a well-nested string has no triple right brackets `${\ggg}$',
then, while moving through this string, the automaton $B_{n,s}$ either rejects,
or works as $B_n$, considering every left bracket as a symbol `${<}$'
of the automaton $B_n$.
Then, since all strings of the form $f_{R_1,\ldots,R_m,\ell_1,\ldots,\ell_{m+1}}g_{i,j,k,m}$
are well-nested and have no triple brackets `${\ggg}$',
Claim~\ref{claim_B_n_on_special_string}
for the automaton $B_n$ and
for strings of the form $f_{R_1,\ldots,R_m}g_{i,j,k,m}$
implies the same claim for the automaton $B_{n,s}$
and for strings of the form $f_{R_1,\ldots,R_m,\ell_1,\ldots,\ell_{m+1}}g_{i,j,k,m}$.

\begin{claim}\label{claim_B_n_s_on_special_string}
Let $m \geqslant 1$ be an integer, let $R_1, R_2, \ldots, R_m \subseteq Q \times Q$
be any non-empty relations, let $\ell_1, \ldots, \ell_{m+1} \in \{0, \ldots, s-1\}$
be indices of left brackets, let $i,j \in Q$ be any two states
and let $k = 1, \ldots,m$ be an integer.
Then the automaton $B_{n,s}$ accepts the string 
$f_{R_1,\ldots,R_m,\ell_1,\ldots,\ell_{m+1}}g_{i,j,k,m}$ 
if and only if $(i,j) \in R_k$.
\end{claim}

For every pair of strings
$f_{R_1,\ldots,R_m,\ell_1,\ldots,\ell_{m+1}}$ and $f_{R'_1,\ldots,R'_m,\ell'_1,\ldots,\ell'_{m+1}}$
with different sequences of non-empty relations, Claim~\ref{claim_B_n_s_on_special_string} allows
one to choose such a continuation $g_{i,j,k,m}$,
that the automaton $B_{n,s}$ will give different answers on the strings
$f_{R_1,\ldots,R_m,\ell_1,\ldots,\ell_{m+1}}g_{i,j,k,m}$ 
and $f_{R'_1,\ldots,R'_m,\ell'_1,\ldots,\ell'_{m+1}}g_{i,j,k,m}$,
accepting one and rejecting the other.
But to distinguish in this way all pairs of strings of the form
$f_{R_1,\ldots,R_m,\ell_1,\ldots,\ell_{m+1}}$,
one also needs to construct such suffixes that will separate any two strings
$f_{R_1,\ldots,R_m,\ell_1,\ldots,\ell_{m+1}}$ and $f_{R'_1,\ldots,R'_m,\ell'_1,\ldots,\ell'_{m+1}}$
with different indices of left brackets in some position: $\ell_k \neq \ell'_k$.

The next goal is to define strings $h_{k,x,m}$, for $m \geqslant 1$, for
a left bracket number $k \in \{1,\ldots, m+1\}$ and
for a bit number $x \in \{0,\ldots,\lfloor\log_2 (s-1)\rfloor\}$,
such that the string $f_{R_1,\ldots,R_m,\ell_1,\ldots,\ell_{m+1}}h_{k,x,m}$
is accepted by the automaton $B_{n,s}$ if and only if $\ell_k[x] = 1$,
that is, if the coefficient at $2^x$ in the binary representation of the number $\ell_k$
equals $1$. Such strings $h_{k,x,m}$ will distinguish any two strings 
$f_{R_1,\ldots,R_m,\ell_1,\ldots,\ell_{m+1}}$ and $f_{R_1,\ldots,R_m,\ell'_1,\ldots,\ell'_{m+1}}$,
in which, for some number $k$, the indices of unmatched left brackets differ: $\ell_k \neq \ell'_k$,
that is, if there is a bit number $x$, such that $\ell_k[x] \neq \ell'_k[x]$.

A string $h_{k,x,m}$, for $m \geqslant 1$, for $k \in \{1,\ldots, m+1\}$ and for 
$x \in \{0,\ldots,\lfloor\log_2 (s-1)\rfloor\}$, is defined as:
\begin{equation*}
h_{k,x,m} = (\#{\gg})^{m-k+1}\#y_{x\bmod  n}{\ggg}y_{\lfloor\frac{x}{n}\rfloor}(\#{\gg})^{k-1}.
\end{equation*}

The next claim is that strings $h_{k,x,m}$ have the desired properties.

\begin{claim}\label{claim_B_n_s_on_special_string_bit_checking}
Let $m \geqslant 1$ be an integer, let $R_1, R_2, \ldots, R_m \subseteq Q \times Q$
be any non-empty relations, let $\ell_1, \ldots \ell_{m+1} \in \{0, \ldots, s-1\}$
be indices of left brackets, let $k \in \{1,\ldots, m+1\}$ be an integer
and let $x \in \{0,\ldots,\lfloor\log_2 (s-1)\rfloor\}$ be a bit number.
Then the automaton $B_{n,s}$ accepts 
the string $f_{R_1,\ldots,R_m,\ell_1,\ldots,\ell_{m+1}}h_{k,x,m}$ 
if and only if $\ell_k[x] = 1$, that is, if the coefficient at $2^x$
in the binary representation of the number $\ell_k$ is $1$.
\end{claim}

\begin{figure}[t]
	\centerline{\includegraphics[scale=1.1]{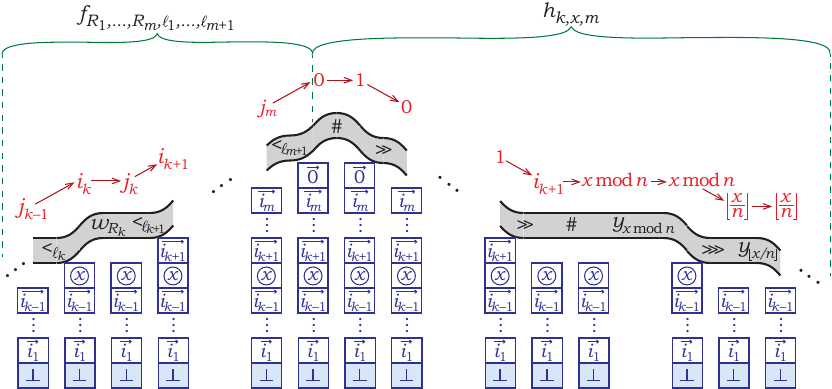}}
	\caption{An accepting computation of the automaton $B_{n,s}$
	on a string $f_{R_1,\ldots,R_m,\ell_1,\ldots,\ell_{m+1}}h_{k,x,m}$, in a case of $\ell_k[x] = 1$.}
	\label{f:idpda_det_manyleft_f_h}
\end{figure}

First, let $\ell_k[x] = 1$. Then there is the following accepting computation
of the automaton $B_{n,s}$ on the string $f_{R_1,\ldots,R_m,\ell_1,\ldots,\ell_{m+1}}h_{k,x,m}$,
illustrated in Figure~\ref{f:idpda_det_manyleft_f_h}.
Since relations $R_1, \ldots, R_m$ are non-empty, one can choose a pair of states of $B_{n,s}$
for each relation: $(i_1,j_1) \in R_1$, \ldots, $(i_m, j_m) \in R_m$.
The automaton reads the string 
$f_{R_1,\ldots,R_m,\ell_1,\ldots,\ell_{m+1}} = {<_{\ell_1}}w_{R_1}{<_{\ell_2}}w_{R_2}\ldots{<_{\ell_m}}w_{R_m}{<_{\ell_{m+1}}}$ as follows:
it begins in the state $j_0 = 0$
and comes to every $t$-th unmatched left bracket `${<}_{\ell_t}$', for $t = 1, \ldots, m$,
in the state $j_{t-1}$,
and then, at each bracket `${<}_{\ell_t}$' it changes its state from $j_{t-1}$ to $i_t$,
reads $w_{R_t}$ changing the state $i_t$ to $j_t$ (this is possible, since $(i_t,j_t) \in R_t$),
and comes to the next unmatched left bracket ${<}_{\ell_{t+1}}$
in the state $j_t$. Finally, the automaton gets to the last unmatched left bracket
of the substring $f_{R_1,\ldots,R_m,\ell_1,\ldots,\ell_{m+1}}$ in the state $j_m$
and leaves the substring in the state $i_{m+1} = 0$.
During this computation,
the automaton pushes symbols $\overrightarrow{i_t}$ onto the stack
at every unmatched left bracket `${<_{\ell_t}}$', for $t \neq k$,
and pushes onto the stack the symbol $\text{\textcircled{$x$}}$
at the $k$-th unmatched left bracket (this can be done, since $\ell_k[x] = 1$).
As a result, the automaton leaves 
the first part of the string $f_{R_1,\ldots,R_m,\ell_1,\ldots,\ell_{m+1}}$
in the state $0$ with the stack contents 
$\alpha = \overrightarrow{i_1} \ldots \overrightarrow{i_{k-1}} \text{\textcircled{$x$}} \overrightarrow{i_{k+1}} \ldots \overrightarrow{i_{m+1}}$.

Now it will be shown, how the automaton $B_{n,s}$ with such a stack
gets through the substring
$h_{k,x,m} = (\#{\gg})^{m-k+1}\#y_{x\bmod  n}{\ggg}y_{\lfloor\frac{x}{n}\rfloor}(\#{\gg})^{k-1}$. The first segment $(\#{\gg})^{m-k+1}$ and the last segment $(\#{\gg})^{k-1}$
are easy to pass: the automaton chooses the state $1$ at symbols `${\#}$'
and does not reject while popping the symbols $\overrightarrow{i_t}$, for $t \neq k$.
The symbol $\text{\textcircled{$x$}}$ is popped out of the stack while reading 
the substring $\#y_{x\bmod  n}{\ggg}y_{\lfloor\frac{x}{n}\rfloor}$.
The automaton can go through this substring as follows: it chooses the state $(x \bmod n)$
at the first symbol `$\#$', passes the substring $y_{x\bmod  n}$,
pops the symbol $\text{\textcircled{$x$}}$ out of the stack 
at the triple right bracket `${\ggg}$' and changes the state from $(x \bmod n)$
to $\lfloor\frac{x}{n}\rfloor$. 
The state $\lfloor\frac{x}{n}\rfloor$ allows the automaton to pass the substring $y_{\lfloor\frac{x}{n}\rfloor}$.
Therefore, the automaton $B_{n,s}$ can move through the second part $h_{k,x,m}$,
entering it in the state $0$ and with stack contents $\alpha$, and accept,
since all states are accepting.
Thus, the automaton accepts the string $f_{R_1,\ldots,R_m,\ell_1,\ldots,\ell_{m+1}}h_{k,x,m}$.

Now let the automaton $B_{n,s}$ accept the string 
$f_{R_1,\ldots,R_m,\ell_1,\ldots,\ell_{m+1}}h_{k,x,m}$, it shall be proved that $\ell_k[x] = 1$.
Consider an arbitrary accepting computation of the automaton $B_{n,s}$
on this string. First, the automaton reads the substring 
$f_{R_1,\ldots,R_m,\ell_1,\ldots,\ell_{m+1}} = {<_{\ell_1}}w_{R_1}{<_{\ell_2}}w_{R_2}\ldots{<_{\ell_m}}w_{R_m}{<_{\ell_{m+1}}}$,
pushing onto the stack some string $\alpha$ of length $m+1$.
It pushes the $k$-th symbol of $\alpha$ onto the stack at the left bracket `${<}_{\ell_k}$',
denote this stack symbol as $c$.
Next, the automaton reads the second part of the string, which is
$h_{k,x,m} = (\#{\gg})^{m-k+1}\#y_{x\bmod  n}{\ggg}y_{\lfloor\frac{x}{n}\rfloor}(\#{\gg})^{k-1}$,
and pops the $k$-th symbol of the string $\alpha$ out of the stack while reading
the substring $\#y_{x\bmod  n}{\ggg}y_{\lfloor\frac{x}{n}\rfloor}$.
Since the automaton rejects at the triple right bracket `${\ggg}$' if it pops a stack symbol
not of the form $\text{\textcircled{$z$}}$, therefore $c = \text{\textcircled{$z$}}$, 
for some $z \in \{0,\ldots,\lfloor\log_2 (s-1)\rfloor\}$.
Since the triple bracket `${\ggg}$' is wrapped in substrings $y_{x\bmod n}$
and $y_{\lfloor\frac{x}{n}\rfloor}$, and since each of them allows only one state,
the automaton comes to the triple bracket `${\ggg}$' in the state $(x \bmod n)$
and leaves this bracket in the state $\lfloor\frac{x}{n}\rfloor$. Such a change of a state
at the triple bracket `${\ggg}$' is possible only if the symbol 
at the top of the stack is $\text{\textcircled{$x$}}$, that is, if $c = \text{\textcircled{$x$}}$.
And the symbol $c$ was pushed onto the stack while reading the left bracket `${<}_{\ell_k}$',
therefore, since $c = \text{\textcircled{$x$}}$, it holds that $\ell_k[x] = 1$.
Claim~\ref{claim_B_n_s_on_special_string_bit_checking} has been proven.

It remains to prove that any deterministic automaton needs a lot of stack symbols 
to simulate the automaton $B_{n,s}$.

\begin{claim}\label{claim_A_simulating_B_n_s_needs_many_stack_alph_symbols}
Let DIDPDA $A$ work over the alphabet $\Sigma$ and recognize the language $L(B_{n,s})$.
Then the automaton $A$ has at least $s(2^{n^2}-1)$ stack symbols.
\end{claim}

Let $N$ be the number of states in the automaton $A$,
and let $M$ be the number of stack symbols it uses.
For the sake of a contradiction, suppose that $M < s(2^{n^2}-1)$.

Then, for a fixed $m$, the automaton $A$ can get at most $N \cdot M^{m+1}$ combinations
of a state and of stack contents after reading strings of the form 
$f_{R_1,\ldots,R_m,\ell_1,\ldots,\ell_{m+1}}$ with $m+1$ unmatched left brackets.
There are exactly $(2^{n^2}-1)^m \cdot s^{m+1}$ strings of the form 
$f_{R_1,\ldots,R_m,\ell_1,\ldots,\ell_{m+1}}$, 
where $R_1, \ldots, R_m \subseteq Q \times Q$ are non-empty relations,
and $\ell_1,\ldots,\ell_{m+1} \in \{0, \ldots, s-1\}$ are left bracket numbers.
Since $M < s(2^{n^2}-1)$,
for values of $m$ large enough, the number $N \cdot M^{m+1}$ is less than $(2^{n^2}-1)^m \cdot s^{m+1}$.
Let $m$ be a sufficiently large integer.
Then, there are two different strings $f_{R_1,\ldots,R_m,\ell_1,\ldots,\ell_{m+1}}$
and $f_{R'_1,\ldots,R'_m,\ell'_1,\ldots,\ell'_{m+1}}$, such that
the deterministic automaton $A$
begins reading both strings in its initial state and leaves both strings in the same state $q$,
with the same stack contents $\alpha$.

Let us first consider the case when the two strings differ in some relation:
$R_k \neq R'_k$, for some $k$. Then there are two states $i,j \in Q$,
such that the pair $(i,j)$ lies in exactly one of the relations $R_k$ and $R'_k$;
without loss of generality, assume that $(i,j) \in R_k$ and $(i,j) \notin R'_k$.
Then, by Claim~\ref{claim_B_n_s_on_special_string},
the nondeterministic automaton $B_{n,s}$ accepts
the string $f_{R_1,\ldots,R_m,\ell_1,\ldots,\ell_{m+1}}g_{i,j,k,m}$
and rejects the string $f_{R'_1,\ldots,R'_m,\ell'_1,\ldots,\ell'_{m+1}}g_{i,j,k,m}$.
Then, the deterministic automaton $A$ does the same, but it cannot
give different answers on these strings,
since on both strings it enters the suffix $g_{i,j,k,m}$ in the state $q$ with a stack $\alpha$.
Therefore, the case of $R_k \neq R'_k$ is impossible.

Then, the two strings $f_{R_1,\ldots,R_m,\ell_1,\ldots,\ell_{m+1}}$
and $f_{R'_1,\ldots,R'_m,\ell'_1,\ldots,\ell'_{m+1}}$ differ in some left bracket numbers:
$\ell_k \neq \ell'_k$. Then one can choose a bit $x$, such that
the coefficients at $2^x$ in numbers $\ell_k$ and $\ell'_k$ differ: $\ell_k[x] \neq \ell'_k[x]$.
Without loss of generality, let $\ell_k[x] = 1$ and $\ell'_k[x] = 0$.
Then, by Claim~\ref{claim_B_n_s_on_special_string_bit_checking},
the automaton $B_{n,s}$ accepts the string $f_{R_1,\ldots,R_m,\ell_1,\ldots,\ell_{m+1}}h_{k,x,m}$
and rejects the string $f_{R'_1,\ldots,R'_m,\ell'_1,\ldots,\ell'_{m+1}}h_{k,x,m}$.
That means that the deterministic automaton $A$ also gives different answers on these strings,
but this is impossible, since on both strings it enters the suffix $h_{k,x,m}$ in the same state $q$
with the same stack $\alpha$.
This is a contradiction, and therefore the automaton $A$
has at least $s(2^{n^2}-1)$ stack symbols.

This finishes the proof of Claim~\ref{claim_A_simulating_B_n_s_needs_many_stack_alph_symbols}
and of Theorem~\ref{theorem_stack_alph_precise_lb_several_opening_brackets}.

It has not been explained yet, 
why Theorem~\ref{theorem_stack_alph_precise_lb_several_opening_brackets}
holds in the degenerate case of only one state: $n = 1$.
If $s = 1$, one can use Theorem~\ref{theorem_stack_alph_precise_lb_1_opening_bracket}.
Since $s \leqslant 2^{n^2}$, the remaining case is $n = 1$ and $s = 2$.
The desired NIDPDA $B_{1,2}$ can be constructed as follows.
An input alphabet is: $\Sigma_0 = \emptyset$,
$\Sigma_{+1} = \{{<}, {\ll}\}$, $\Sigma_{-1} = \{{>},{\gg}\}$.
At the single left bracket `${<}$' the automaton pushes the symbol $0$ onto the stack,
at the double left bracket `${\ll}$' it pushes $1$.
At the single right bracket `$>$'
the automaton $B_{1,2}$ moves forward if there is $0$ at the top of the stack,
and rejects, if there is $1$; and works vice versa at the double right bracket `${\gg}$'.
The automaton $B_{1,2}$ thus defined accepts a string if and only if
each single left bracket `${<}$' corresponds to a single right bracket `${>}$',
and every double left bracket `${\ll}$' corresponds to a double bracket `${\gg}$'.

Let $A$ be a deterministic automaton recognizing the language $L(B_{1,2})$.
Then, it has at least $2$ states: an accepting state to accept the empty string, and
a rejecting one to reject the string ${<}{\gg}$.
It shall be proved that the automaton $A$ has at least $2(2^{1^2}-1) = 2$ stack symbols.
Assume the opposite, that there is only one stack symbol in the stack alphabet of $A$.
There are two different strings $u$ and $u'$
of the same length over an alphabet $\{{<},{\ll}\}$, such that the automaton $A$ finishes reading them in the same state $q$.
Consider the string $v$, that matches the string $u$ with right brackets of the same types,
so that $B_{1,2}$ accepts the string $uv$. On the other hand, since $u \neq u'$,
there is a left bracket in $u'$ that differs from the left bracket at the same position
in $u$, and therefore this bracket in $u'$ does not correspond to the matching right bracket in $v$,
and the automaton $B_{1,2}$ rejects the string $u'v$. Then, the automaton $A$
also accepts the string $uv$ and rejects the string $u'v$,
but it leaves the substrings $u$ and $u'$ in the same state $q$ with the same stack contents,
this is a contradiction.
\end{proof}

Note that the size of the stack alphabet
in Theorem~\ref{theorem_stack_alph_precise_lb_several_opening_brackets}
of $B_{n,s}$ is linear in the number of states
and is logarithmic in the number of left brackets. 
The theorem is stated for a number of left brackets not exceeding $2^{n^2}$.
As a matter of fact, this lower bound cannot possibly be extended for an unbounded number of brackets,
because an $n$-state nondeterministic automaton
has only $2^{|\Gamma|n^2}$ possible transition functions by left brackets,
and the simulating automaton could remember this function instead of a bracket.
Thus, the restriction on the number of left brackets in the theorem is close to the best possible.

\section{Conclusion}

There are some open questions on the state complexity of input-driven pushdown automata.
The lower bound on the number of states needed for determinization is precise, 
the input alphabet is fixed and the stack alphabet is two-symbol,
so it is hard to improve anything here.
However, in the lower bound on the number of stack symbols in a deterministic automaton
obtained by NIDPDA determinization,
witness automata use
a stack alphabet linear in $n$.
Is it possible to obtain the same lower bound
using a stack alphabet smaller than linear in $n$?

Jir\'askov\'a and Okhotin~\cite{JiraskovaOkhotin_idpda_sc}
investigated the state complexity of operations on DIDPDA.
Lower and upper bounds on the number of states for concatenation and for the Kleene star
are asymptotically tight: $mn^n$ and $m(n^n+2^n)$ for concatenation of automata
with $m$ and $n$ states, and $n^n$ and $n^n+2^n+1$ for the Kleene star.
One can try to find the exact state complexity.

Ogawa and Okhotin~\cite{OgawaOkhotin} established asymptotically precise bounds
for determinization of event-clock input-driven pushdown automata,
but one can try to make these bounds even more precise.

Rose and Okhotin~\cite{RoseOkhotin} 
introduced probabilistic input-driven pushdown automata,
determinized them and proved the asymptotically precise lower bound on the number of states
needed for determinization. It remains an open problem to prove a lower bound on the
number of stack symbols in a deterministic automaton needed for determinization
of probabilistic input-driven pushdown automata.

\section*{Acknowledgements}

I am grateful to Alexander Okhotin for his advices on the presentation
and for help in translating the paper to English.

This work was supported by
the Ministry of Science and Higher Education of the Russian Federation,
agreement 075-15-2022-287.

\end{document}